\documentclass{article}
\usepackage{spconf,amsmath,graphicx}
\usepackage{booktabs}
\usepackage{hyperref}
\usepackage{dsfont}
\usepackage{amsmath,amssymb}
\usepackage{array}

\title{A Framework for  Generative and Contrastive Learning of Audio Representations}
\name{Prateek Verma, Julius Smith}
\address{Center for Computer Research in Music and Acoustics, Stanford University\\
{\tt prateekv@ccrma.stanford.edu,jos@ccrma.stanford.edu}
}

\begin{document}
%
\maketitle
\begin{abstract}
In this paper, we present a framework for contrastive learning for
audio representations, in a self supervised framework without access
to any ground truth labels. The core idea in self supervised
contrastive learning is to map an audio signal and its various
augmented versions (representative of salient aspects of audio like
pitch, timbre etc.) to a space where they are close together, and are
separated from other different signals. In addition we also explore
generative models based on state of the art transformer based
architectures for learning latent spaces for audio signals, without
access to any labels. Here, we map audio signals on a smaller scale to
discrete dictionary elements and train transformers to predict the
next dictionary element. We only use data as a method of supervision,
bypassing the need of labels needed to act as a supervision for
training the deep neural networks. We then use a linear classifier
head in order to evaluate the performance of our models, for both self
supervised contrastive and generative transformer based representations that are
learned. Our system achieves considerable performance, compared to a fully
supervised method, with access to ground truth labels to train the
neural network model. These representations, with availability of
large scale audio data show promise in various tasks for audio
understanding tasks.\footnote{This work was done when Prateek Verma
  was a Research Assistant with Julius Smith at Center for Computer
  Research in Music and Acoustics, at Stanford University in Spring
  2020}
\end{abstract}
%

\section{Introduction and Related Work}
\label{sec:introduction}
The second wave of the development of neural network interest began with
the paper by Hinton et al.{\ }\cite{hinton2006reducing} in 2006, which
showed the ability of neural nets to successfully compress images into
a meaningful latent representation. There has been a series of recent
works with the advent of deep learning and neural networks to
represent the signals of interest, viz., images \cite{chen2020simple}, text \cite{Mikolov2013-w2v}
or audio \cite{oord2018representation, haque2019audio}, by a vector representation
that is suitable for the task of interest. Broadly, these ideas can be
classified into generative \cite {radford2015unsupervised} or
discriminative methods \cite{doersch2017multi}. Within these two
there can again be approaches pertaining to supervised and self
supervised approaches, which amount to using labels of interest to
guide the neural network to learn the representation via the final
layers \cite{Vggish,verma2019neuralogram} or intermediate layers
\cite{verma2018neural} for a particular application. For supervised
approaches, the latent spaces are often used to guide one modality
into another. As an example, SoundNet \cite {aytar2016soundnet} used
latent spaces of images as a target to guide a deep neural network to
learn from waveforms. Due to the availability of large amounts of
unlabeled data, such approaches were able to scale and outperform
state of the art acoustic understanding works existing at the time.
This is mainly because of the size of the dataset used for training
\cite{ellis1999size}, and the contents of the latent code
encapsulating much richer information than one-hot raw labels. The main motivation for the current work,
inspired from similar ideas in computer vision \cite{chen2020simple}
and natural language processing \cite{sentimentneuron, devlin2018bert},
is to use the data itself for self-learning or supervision,
which we call in this case a self-supervised learning setup. In
natural language processing, models such as BERT \cite{devlin2018bert}
use masked inputs to predict the entities that were masked, yielding
better representation, and solving several downstream tasks.
Such ideas have been proposed
for audio \cite{chuang2019speechbert} have shown promising results. Contrastive
based loss functions have been proposed in various supervised setups
for audio and music understanding \cite{jansen2020coincidence}.

\begin{figure*}[t]
    \centering
    \includegraphics[width=18cm,height=4cm]{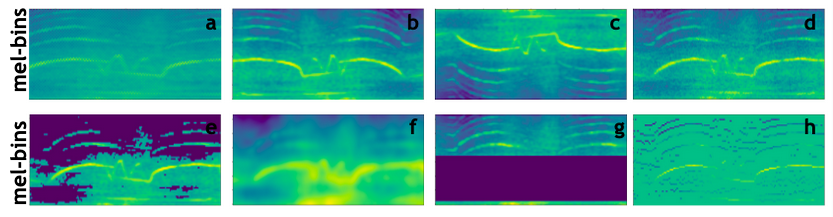}
    \caption{\textbf{Various Spectral Data Augmentation Representations} a) adding checkerboard noise b) time reversal c) flipping d) adding noise e) energy based thresholding f) spectral envelope based representation per spectral slice g) band-stop filtering h) retaining salient spectral peaks. For each of the representations, there can be multiple level of distortions}
    \label{fig:method}
\end{figure*}
In our first approach, we map augmented versions of the audio signal to
the same space while pushing them away from augmented versions of different
audio signals. This follows the approach first proposed in
\cite{chen2020simple} and subsequent works. In the generative
representation learning method, we draw on advances in representation
learning \cite{van2017neural} along with the ability of transformer architectures
\cite{vaswani2017attention} to understand dependencies
across time. We propose to learn transformers operating on a learned dictionary
of latent space elements using simple clustering algorithms like
k-means, which is a simple yet powerful
variant of vector-quantized variational auto-encoders
\cite{van2017neural}. By learning to predict accurately the next
elements of this learned dictionary space, the hypothesis is that we
would learn a good representation of the signal, encapsulating what
has already happened in the past. Similar ideas have been proposed in
audio problems such as packet loss concealment using deep neural networks \cite{verma2020deep}. Contributions of the current work are:
\begin{itemize}
\item[i)] We propose a framework for training a deep neural network
  encoder (a convolutional network), using a contrastive loss, purely
  from data and its augmented variants, for audio understanding. The
  proposed data augmentation techniques, derived from signal
  processing, are designed to capture fundamental aspects of audio.
\item[ii)] We also propose a dictionary-based generative model based
  on transformers for learning an audio representation, which can then
  be used along with a linear head for downstream audio tasks.
\end{itemize}

\section{Dataset and Audio Representation}
We choose to work with a standard acoustic scene analysis dataset UrbanSound 8K \cite{salamon2014dataset}. This dataset contains 8732 labeled sound excerpts (less than 4s) of urban sounds from 10 classes: air conditioner, car horn, children playing, dog bark, drilling, engine idling, gun shot, jackhammer, siren, and street music. We use the folds given along with the meta-data information for all of our experiments. For audio signals not equal to 4s we append them with repeating the signal to make all of the signals 4s in duration to have consistency across the dataset. We choose to represent audio signals as log magnitude mel spectrograms,  which have been used in the literature \cite{Vggish}. The audio is downsampled to 16000Hz and converted to its log-magnitude mel-spectral representation with a 10ms hop size, 30ms window, 1024-point FFT, mapped to 128 bins, yielding a 128x400 matrix representation for every audio excerpt present in the dataset, using the librosa library \cite{mcfee2015librosa}. Without loss of significant resolution, we chose to down-sample from this representation by a factor of 2 to get an interpolated version of the mel-spectrogram at size 64x200, in order to save computation time, without drastically removing the contents of the signal.

\section{Method}\label{sec:methods} 
In this section, we explain how we use two distinct approaches to learn audio representations in a self-supervised manner. The two methods, (1) using a framework for contrastive learning and (2) temporal learning based on transformers, are described in the following sections. There are no annotated labels for training purposes---the data itself is used for supervision. Contrastive learning methods, which follow work in vision, combine ideas from traditional signal processing in order to create multiple feature-specific input representations (augmentations) that are then mapped to a common space. In addition to this approach, we also propose the idea of learning a feature representation by predicting what is going to happen next. If a feature vector can encapsulate all the contents of what has happened in the past, then it becomes a sufficient statistic, and can predict the future. In other words, we use future step prediction as a pseudo-task for learning a representation for the contents of the audio.
\subsection{Contrastive Learning}\label{contrastive learning} 
\subsubsection{Data-Augmentation}\label{data_augmentation}
As described in the previous section, the framework proposed by Hinton
et al.{\ }\cite{chen2020simple} learns to map the embeddings of different
augmentations of the input signal to the same space, and that of
different audio signals to different spaces. We have proposed several
data augmentation techniques adapted for audio signals. The role of
data augmentation in helping to generalize deep neural networks has
been widely studied, particularly for audio signals
\cite{mcfee2015software, schluter2015exploring}. A point to note is
that, as the training set keeps getting larger through augmentation, the size and diversity
of the data often become sufficient to achieve state of the art results
\cite{ellis1999size}, often overcoming model limitations. For this reason,
unsupervised approaches of this nature are very promising. Given mel-spectral representation, we do in total 9 augmentation strategies. i) Adding checkerboard noise: We randomly turn off alternative time-frequency bins, by a spacing factor of 2,3,4 ii) we time reverse and iii) flip upside down mel-spectra iv) do Amplitude scaling in range of +/- 10db v) band-limiting the mel-spectra with random onset and the width of the filter, shared across all the time-bins vi) retaining only spectral peaks vii) retaining t-f bins above certain energy threshold viii) adding random noise ix) extracting the spectral envelop and throwing away spectral peaks.

\begin{figure}[t]
    \centering
    \includegraphics[width=1.0\linewidth]{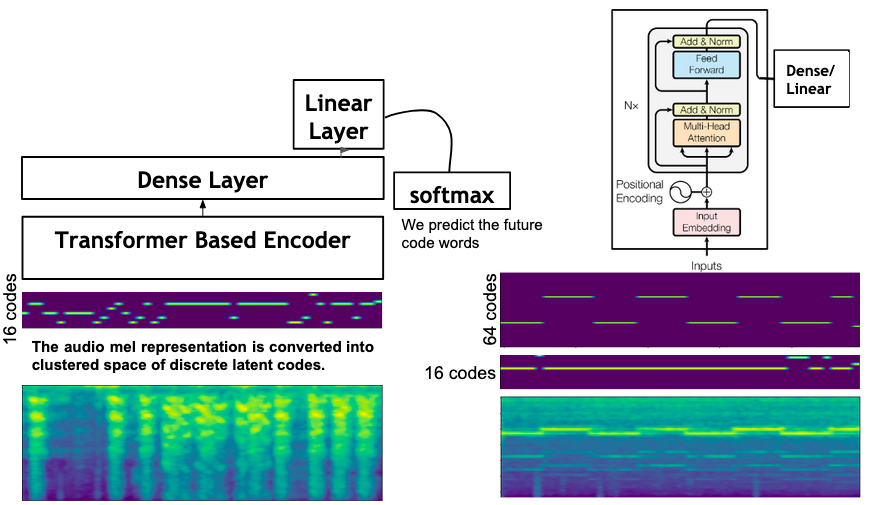}
    \caption{\textbf{Block diagram of our method.} 
      Transformer predicts the next code, and dense layer acts as a representation of the audio, Choosing smaller number of clusters, makes all of the sinusoidal sounds get assigned to a single
      cluster. Increase in the number of clusters, we see individual
      frequency components are learned as separate elements of the
      dictionary. Transformer block taken from \cite{vaswani2017attention}}
    \label{fig:method}
\end{figure}
\subsubsection{Algorithm} The algorithm follows the framework proposed by Hinton et. al \cite{chen2020simple}. We fix our encoders to be a 16-layer VGG neural network architecture (choice to save on computation time), followed by a dense layer which converts the spectral representation to a vector of size 512. Let $X_i$ and $X_j$ be sampled from the same mel-spectral representation $X$, where $i$ and $j$ can be two arbitrary data augmentations sampled from the above mentioned strategies. A fixed convolutional encoder $f_{conv}$, VGG-16 network in this case, transforms the spectral representations to a feature space $e_i$ and $e_j$ respectively. In our case, both are output of a dense layer of size 512 followed after the convolutional layers of the VGG architecture. A 2-layer MLP network, (with 128 neurons each) converts these to a more separable space $z_i$ and $z_j$. For every batch of size $N$, we now create a new batch of size $2*N$, such that the consecutive elements ($2k,2k+1$) are from the same augmented versions of a single mel-spectra. We now compute a distance matrix $s_{i,j}$, of the output vectors of the MLP network, where $s_{i,j} = z^{T}_{i}  z_{j}/||z_{i}||.||z_{j}||$. The loss function which is minimized is negative cross entropy loss which is defined below, with $\tau$ being the temperature parameter, 

\begin{equation}
    l(i,j) = - \log  \frac{\exp(s_{i,j}/\tau)}{\sum_{k=1}^{2N} \mathds{1}_{k\neq i} \exp(s_{i,j}/\tau)}
\end{equation}

and the total which we minimize is, \begin{equation}
    \mathcal{L} = \frac{1}{2N} \sum_{k=1}^{N}[ l(2k-1,2k)+ l(2k,2k-1)]
\end{equation} Intuitively, we in a way, take softmax over rows of the distance matrix, and try to make similar augmented versions of the same spectra closer. We tweak with various hyper-parameters with respect to this setup, choosing $\tau$ from [0.1, 0.5, 1] and batch size [64, 128, 256].  
\subsection {Generative learning using Transformers} \label{generative learning}
In this approach, we combine the strengths of two unsupervised
learning approaches, viz., k-means clustering and transformer-based
dependency learning. The goal here is to learn a representation of an
audio signal using self supervision without access to labels during training of the network.

\subsubsection{Latent space learning}
We deploy a two-step approach for this setup. We convert audio snippets
into discrete latent spaces. For the audio input consisting of log-magnitude,
mel-scale spectrogram input (64 spectral bins and 200 frame times),
we convert the audio signal into patches of width
4 frames, which is a design choice. The goal is to learn, and convert the
input mel spectrogram according to a discrete latent code based
dictionary. VQ-VAE \cite{van2017neural} is a variant of vector
quantization, the simplest one from the family, being k-means clustering. There has
been previous work on converting audio into latent-space based meaningful
representation \cite{verma2019neuralogram}. We take
spectrogram patches of 4 spectral frames across frequency, and train auto-encoders taking
input as 64x4 (defined henceforth patch of audio spectra), with three-layer fully connected encoder and decoders with
1024 neurons in each layer, and the size of the bottleneck layer set to
16. For all of the dataset, we convert the mel-spectrogram representation
into a discrete latent code based on the cluster to which the current
embedding of the audio patch belongs after k-means
clustering. k-means clustering is carried out with all of the audio
patches present in the training set, and yields the corresponding
cluster index for the current patch depending on the number of clusters
we choose. With above specifications, we convert the mel-spectrogram into a time series of 50 elements (four-frame audio patches), where each element corresponds to the cluster to which it belongs to (with possible cluster assignments as 16/64/256).

\subsubsection{Transformers on discrete latent spaces} Transformers \cite{vaswani2017attention} have revolutionized the field of natural language processing in achieving state of art results in language understanding, and synthesis and many more. The ability of these models to understand context much better than RNNs, dilated convolutions has been the reason we choose to work with them. A more detailed description is mentioned in \cite{vaswani2017attention}, and is beyond the scope of the current work. Briefly, they use multiple attention heads in order to understand salient features, and use feed-forward nets to transform the features it is attending to a more appropriate latent space depending on the problem of interest. For our problem, we draw inspiration from the work for sentiment neuron \cite{sentimentneuron} and other generative models like BERT \cite{chuang2019speechbert,devlin2018bert}. If we can predict what is going to happen next given the previous context, then we can encapsulate the contents of the audio signal. This vector is used for classification task of interest using a linear head. \par
The exploration space is vast (number of attention heads, Transformer layers, size of the dictionary etc), and with the availability of limited computing resources, we come up with the following
experimental setup.  We predict the next time step from the previous input representation. We take context as half of the input signal i.e. 25 discrete points in our codebook corresponding to the mel-spectral patches, with the number of clusters (vocab) being 256.
The model consists of 128 dimensional feed forward neural
network, with 3 layers as an encoder in each Transformer module, with
8 attention heads and the size of our embedding fixed as 32. A Dense layer converts the output embedding to a smaller 16-dimensional representation. For regularization, for 20\% of the training steps, we corrupt the inputs by a random factor deciding how many inputs to corrupt for a particular training step. In order to predict the next code, we predict a 256 dimensional output (corresponding to one hot vector of next code), from the flattened output of last layer of the transformer followed by a softmax layer, to minimize the cross entropy loss between the desired one hot label, and the logits predicted by the network. The training was carried out for 50 epochs using
a learning rate of 1e-4, and then decaying it by a factor of 10, every
20 epochs using Adam optimizer, using V100 GPUs in TensorFlow framework in the Google Cloud environment.

\section{Results and Discussion}\label{sec:evaluation}

\begin{figure}[t]
    \centering
    \includegraphics[width=1.0\linewidth]{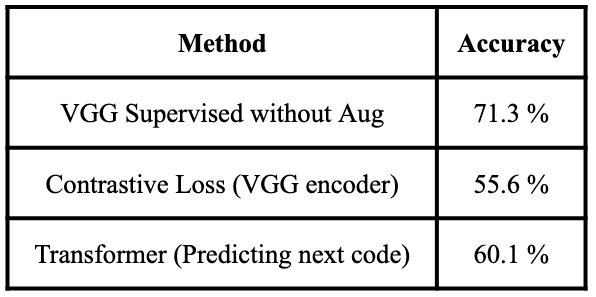}
    \caption{\textbf{Results of our experiments} 
     These are the results obtained after a preliminary parameter search for optimal batch size and temperature $\tau$, for contrastive loss based approach, and number of attention heads, dimensions of feed forward network, latent code etc, for Transformers }
    \label{fig:method}
\end{figure}
We compare the best performing generative models, with the set-up we described in the previous section for contrastive loss and a baseline system.
Considerable performance improvements can happen with having state of the art encoder models like ResNet-101, DenseNet. As we see, we achieve a comparable gap between supervised and unsupervised approaches (15\% in our case vs around 10\% achieved in \cite{chen2020simple}. We believe this is indeed a promising direction, and with much larger exploration of hyper-parameters we can narrow this gap in future (optimal choice of latent code size, $\tau$ etc. Our work shows viability of contrastive and generative based approaches for audio understanding. Since we do not train our models on annotated data, the model performance can show considerable gains with pre-training on large amounts of dataset e.g \cite{Vggish}, as performance often improves with more training data. \cite{ellis1999size}. In this study, we compared two methods with almost the same amount of training data in a controlled experimental setting. Further, these results also show the power of generative models, in achieving strong results, with a fraction of parameters that are used in contrastive based approach's. The idea of combining clustering based dictionary learning approach with that of Transformer is a powerful concept that has many applications ahead.

\section{Conclusion and Future Work}\label{sec:conclusion}
We have proposed a method based on two distinct approaches: i)
contrastive learning based on data augmentation, and ii) transformer-based
next-element prediction for audio representation learning. We show
how our method achieves considerable performance purely by training
the deep neural networks on the data, and using a linear head, (as
opposed to multilayer nonlinear heads with labels) to classify the
learned representations for various applications. The strengths of
contrastive and transformer-based representation learning methods
suggest various applications in problems where large scale labeling
is not feasible. We can envision various modifications of the existing
algorithm such as different loss functions, using state of the art encoders, and combining these ideas with clustering and multitask learning for
achieving even better performance, purely in a self supervised
setup. These self-supervised learned embeddings can further be
combined with other neural architectures for various applications in
vision, speech, and natural language processing.

\bibliographystyle{IEEEbib}
\raggedright
\bibliography{bib}

\end{document}